\documentclass[10pt, twocolumn, pre, aps, superscriptaddress, showpacs]{revtex4-1}
\usepackage{amsmath, graphicx, subfigure, tikz}
\begin{document}

\title{Axial, Planar-Diagonal, Body-Diagonal Fields on the Cubic-Spin Spin Glass in d=3:\\
A Plethora of Ordered Phases under Finite Fields}

\author{E. Can Artun}
    \affiliation{T\"UBITAK Research Institute for Fundamental Sciences, Gebze, Kocaeli 41470, Turkey}
    \affiliation{Faculty of Engineering and Natural Sciences, Kadir Has University, Cibali, Istanbul 34083, Turkey}
\author{Deniz Sarman}
    \affiliation{Department of Physics, Bo\u{g}azi\c{c}i University, Bebek, Istanbul 34342, Turkey}
\author{A. Nihat Berker}
    \affiliation{Faculty of Engineering and Natural Sciences, Kadir Has University, Cibali, Istanbul 34083, Turkey}
    \affiliation{T\"UBITAK Research Institute for Fundamental Sciences, Gebze, Kocaeli 41470, Turkey}
    \affiliation{Department of Physics, Massachusetts Institute of Technology, Cambridge, Massachusetts 02139, USA}

\begin{abstract}
A nematic phase, previously seen in the $d=3$ classical Heisenberg spin-glass system, occurs in the n-component cubic-spin spin-glass system, between the low-temperature spin-glass phase and the high-temperature disordered phase, for number of spin components $n\geq 3$, in spatial dimension $d=3$, thus constituting a liquid-crystal phase in a dirty (quenched-disordered) magnet. Furthermore, under application of a variety of uniform magnetic fields, a veritable plethora of phases are found. Under uniform magnetic fields, 17 different phases and two spin-glass phase diagram topologies (meaning the occurrences and relative positions of the many phases), qualitatively different from the conventional spin-glass phase diagram topology, are seen. The chaotic rescaling behaviors and their Lyapunov exponents are calculated in each of these spin-glass phase diagram topologies.  These results are obtained from renormalization-group calculations that are exact on the $d=3$ hierarchical lattice and, equivalently, approximate on the cubic spatial lattice. Axial, planar-diagonal, or body-diagonal finite-strength uniform fields are applied to $n=2$ and 3 component cubic-spin spin-glass systems in $d=3$.

\end{abstract}
\maketitle

\begin{figure}[ht!]
\centering
\includegraphics[scale=0.55]{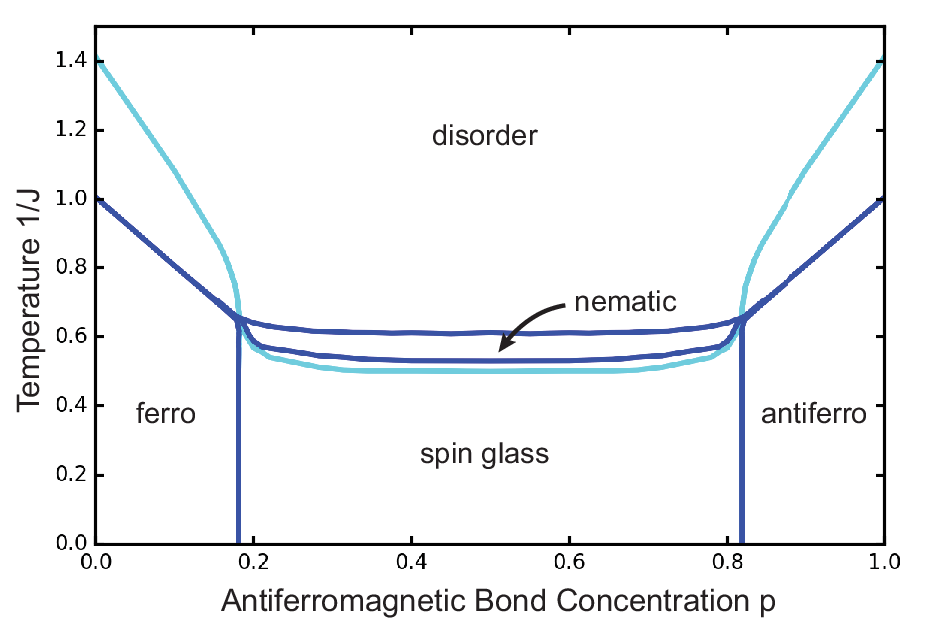}
\caption{Calculated \cite{Artun} phase diagrams for the cubic-spin spin-glass systems for zero external field in spatial dimension $d=3$. The phase diagrams are, from top to bottom, for number of spin components $n=2$ (green lines) and $n=3$ (blue lines).  The vertical ferromagnetic-spinglass-antiferromagnetic phase boundaries overlap for $n=2$ and 3.  A nematic phase appears for $n=3$, between the low-temperature spin-glass phase and the high-temperature disordered phase. The calculation used in the current study yields this phase diagram at zero field.}
\end{figure}

\section{Cubic-Spin Spin-Glass System and Nematic Phase in a Dirty Magnet}

Spin-glass systems have an inherent quantifiable chaos under scale change \cite{McKayChaos,McKayChaos2,BerkerMcKay,McKayChaos4,ZZhu,Katzgraber3,Fernandez,Fernandez2,Eldan,Wang2,Parisi3} and thus provide a universal classification and clustering scheme for complex phenomena \cite{classif}, as well as rich ordering phenomena such as spin-glass sponge ordering \cite{sponge} with separate interior or exterior chaos.  Spin-glass studies have been done overwhelmingly with Ising $s_i=\pm1$ spins.  However, a recent study \cite{Tunca} with classical Heisenberg spins $\vec s_i$ that can continuously point in $4\pi$ steradians found, instead of spin-glass order, nematic order, meaning a liquid-crystal phase in a dirty magnet.  Furthermore, cubic-spin spin-glass systems have yielded both the nematic phase and the spin-glass phase, in the same phase diagram.\cite{Artun}

Recalling the phases of a conventional Ising spin-glass phase diagram, the application of a uniform magnetic field to the antiferromagetic system extends the antiferromagnetic phase in the magnetic field direction, yielding a concrete phase diagram. The application of even an infinitesimal uniform magnetic field to the ferromagnetic phase, destroys the ferromagnetic phase. It has been calculated \cite{Berker0} that the application of even an infinitesimal uniform field to an Ising $(n=1)$ spin-glass phase destroys the spin-glass phase.  The situation is dramatically different, for cubic $(n\geq 2)$ spin systems, as we see below.

Thus, our aim in this article has been to investigate two interesting phenomena of the spin-glass system, namely the replacement of the spin-glass phase by the nematic phase in the continuum-spin Heisenberg spin-glass \cite{Tunca} and the occurrence of both a spin-glass phase and a nematic phase in the discrete yet spin-multidimensional cubic-spin spin-glass \cite{Artun}.  The effect of application of a variety of magnetic fields in a variety of spin dimensions thus yields phase diagrams with a large number of very different phases.

For an $n$-component spin system, $n$ different types of magnetic fields can be applied, each type with $n'\leq n$ magnetic-field components, and qualitatively different effects, as seen below.  In this study, we perform a global renormalization-group study for $n=2$ and 3-component cubic-spin spin-glass systems, in turn applying axial $(n'=1)$, planar-diagonal $(n'=2)$, and body-diagonal $(n'=3)$ magnetic fields,
yielding 16 different phases and two spin-glass phase diagram topologies different from the conventional spin-glass phase diagram topology.

The ordered phases include phases which are axial (single spin direction) $x$ aligned, planar (two spin directions) $yx$ aligned, body (three spin directions) $xyz$ aligned; coexisting $x$ or $y$ aligned, coexisting $x$ or $y$ or $z$ aligned; $x$ and $y$ alternating aligned; $x$ antiferromagnetic, coexisting $x$ or $y$ antiferromagnetic, coexisting $x$ or $y$ or $z$ antiferromagnetic; field-continued (from zero field) and field-induced (at finite field) spin-glass phases; in a rich variety of phase diagram topologies (meaning the occurrences and relative positions of the many phases).

\begin{figure*}[ht!]
\centering
\includegraphics[scale=0.37]{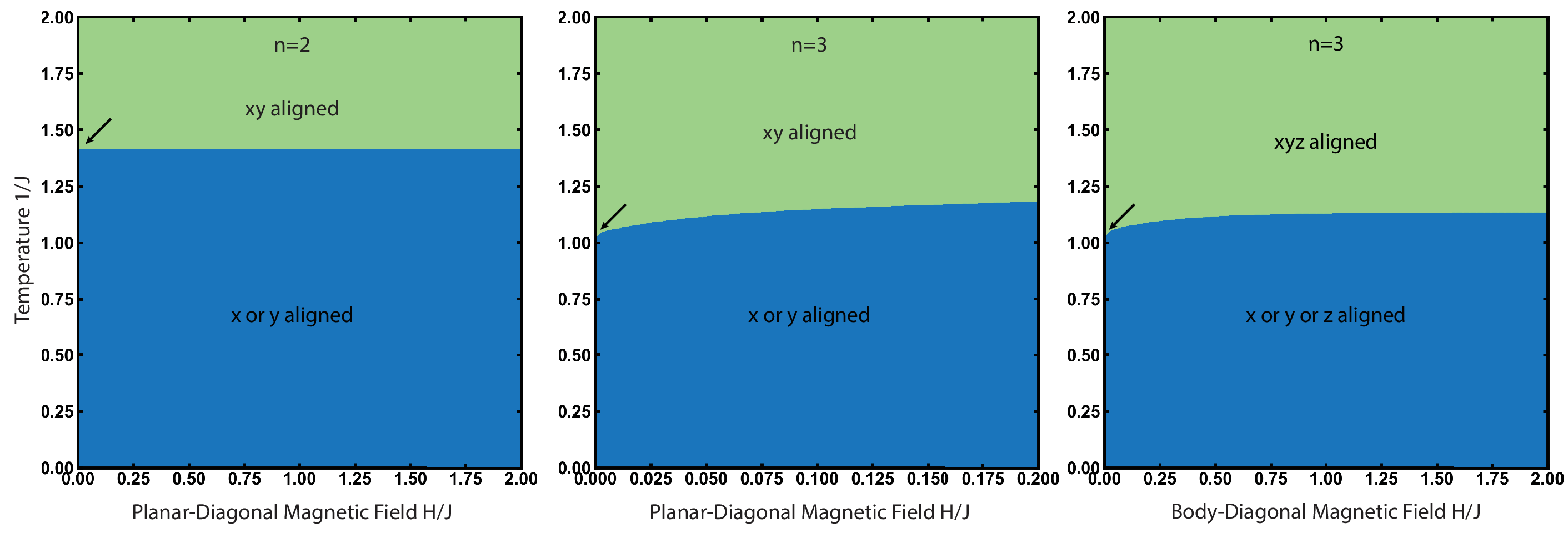}
\caption{Calculated phase diagrams for the cubic-spin system under planar-diagonal (in the $+xy$ direction) and body-diagonal (in the $+xyz$ direction) magnetic fields, in spatial dimension $d=3$, for $p=0$, namely the ferromagnetic system. In all three cases, a uniaxially aligned symmetry-broken ($x$ or $y$ or $z$ aligned) ordered phase occurs at low temperatures and persists to all high fields.  A phase aligned along the applied magnetic field occurs at high temperatures.  The phase transition temperatures between these two phases are essentially independent of field strength and join, at the left intercept of the panels, the ferromagnetic transition temperature (marked by arrow) of the zero-field systems. The ordered phases at finite-field are doubly (in planar-diagonal field) or triply (in body-diagonal field) degenerate. These degeneracies double at ordered phases joined at zero field. The renormalization-group sinks of each phase, to which all points of the phase map under renormalization-group, are given in Table I.  These sinks epitomize the ordering of their respective phases.}
\end{figure*}

\section{Model and Method}

The $n$-component cubic-spin spin-glass system in an $n'$-component uniform magnetic field is defined by the Hamiltonian, where $\beta=1/kT$,
\begin{equation}
-\beta \mathcal{H}=\sum_{\langle ij \rangle} [J_{ij} \vec s_i \cdot \vec s_j +  \vec H \cdot (\vec s_i + \vec s_j)] \equiv \sum_{\langle ij \rangle} -\beta {\cal H}_{ij} \,,
\end{equation}
where $\vec s_i$ can be in $2n$ different states $\pm \hat{u}$ at each site $i$, $\hat{u}$ being orthogonal unit vectors in spin space. The $n'$-component uniform magnetic field is $\vec H = H(\hat{u}_1 +...+ \hat{u}_{n'})$, with of course $n'\leq n$.  The sum is over nearest-neighbor pairs of site $\langle ij \rangle$. The interaction $J_{ij}$ is ferromagnetic $+J>0$ or antiferromagnetic $-J$ with probabilities $1-p$ and $p$, respectively.

The hierarchical-lattice \cite{BerkerOstlund,Kaufman1,Kaufman2} exact renormalization-group solution or, equivalently, the Migdal-Kadanoff \cite{Migdal,Kadanoff} approximate renormalization-group solution of such system has been described in detail.  The construction of the hierarchical lattice, to be solved exactly, is by first constructing strands of $b$ nearest-neighbor interactions $-\beta \mathcal{H}_{ij}$ in series.  Here $b=3$ is the length rescaling factor.  Then $b^{d-1}$ such strands are connected in parallel.  Here $d=3$ is the spatial dimensionality. The hierarchical lattice is obtained by self-imbedding this graph infinitely.  The renormalization-group solution is effected by proceeding in the reverse direction. Alternately, and algebraically equivalently, the Migdal-Kadanoff approximation is constructed by rendering the cubic system renormalizable by bond removing, then reducing via decimating $b$ interactions in series to a single interaction, and then by adding $b^{d-1}$ such interactions to compensate for the bond removing. The hierarchical-lattice realization makes the physically intuitive, much-used Migdal-Kadanoff approximation a realizable, therefore robust, approximation, as has been used in turbulence \cite{Kraichnan}, electronic systems \cite{Lloyd}, and polymers \cite{Flory,Kaufman}. For recent works using hierarchical lattices, see \cite{Clark,Kotorowicz,ZhangQiao,Jiang,Chio,Myshlyavtsev,Derevyagin,Shrock,Monthus,Sariyer}.

The transformation is best implemented algebraically by writing the exponentiated nearest-neighbor Hamiltonian, namely the transfer matrix between two neighboring sites, namely
\begin{multline}
\textbf{T}_{ij} \equiv e^{-\beta {\cal H}_{ij}} = e^{J \vec s_i \cdot \vec s_j + \vec H \cdot (\vec s_i + \vec s_j)} = \\
\left(
\begin{array}{cccccc}
e^{J+2H} & e^{-J} & e^H & e^H & e^H & e^H\\
e^{-J} & e^{J-2H} & e^{-H} & e^{-H} & e^{-H} & e^{-H}\\\textbf{}
e^H & e^{-H} & e^J & e^{-J}& 1 & 1 \\
e^H & e^{-H} & e^{-J} & e^J & 1 & 1 \\
e^H & e^{-H} & 1 & 1 & e^J & e^{-J} \\
e^H & e^{-H} & 1 & 1 & e^{-J} & e^J \end{array} \right),\\
\left(
\begin{array}{cccccc}
e^{J+2H} & e^{-J} & e^{2H} & 1 & e^H & e^H\\
e^{-J} & e^{J-2H} & 1 & e^{-2H} & e^{-H} & e^{-H}\\
e^{2H} & 1 & e^{J+2H} & e^{-J}& e^H & e^H \\
1 & e^{-2H} & e^{-J} & e^{J-2H}& e^{-H} & e^{-H}\\
e^H & e^{-H} &  e^H & e^{-H} & e^J & e^{-J} \\
e^H & e^{-H} &  e^H & e^{-H} & e^{-J} & e^J \end{array} \right),\\
\left(
\begin{array}{cccccc}
e^{J+2H} & e^{-J} & e^{2H} & 1 & e^{2H} & 1\\
e^{-J} & e^{J-2H} & 1 & e^{-2H} &  1 & e^{-2H}\\
e^{2H} & 1 & e^{J+2H} & e^{-J}& e^{2H} & 1\\
1 & e^{-2H} & e^{-J} & e^{J-2H}&  1 & e^{-2H}\\
 e^{2H} & 1&  e^{2H} & 1 & e^{J+2H} & e^{-J} \\
 1 & e^{-2H}&  1 & e^{-2H} & e^{-J} &e^{J-2H} \end{array} \right),
\end{multline}
respectively for axial, planar-diagonal, and body-diagonal magnetic fields, respectively corresponding to magnetic field component numbers $n'=1,2,3$.  In the transfer matrices in Eq. (2), the consecutive states are the plus and minus directions of the sequence of spin-space unit vectors $\hat{u}$.  The transfer matrix is given here for number of spin components $n=3$.  The generalization to arbitrary $n$ is obvious, and used in this work. These are the transfer matrices of the initial points of renormalization-group trajectories.  The transfer matrices of course do not conserve these forms under the  renormalization-group transformations that make up the renormalization-group trajectory, but evolve to one of the 17 asymptotic forms (sinks) given in Table I and Fig. 5 below, thereby determining the ordering and thermodynamic phase of the initial point of the renormalization-group trajectory.

For quenched random systems such as here, 5,000 graphs, defined after Eq. (1) above, are created by randomly choosing $+J$ or $-J$.  The renormalization-group solution proceeds by randomly associating $b^d$ such graphs, to generate the renormalized 5,000 graphs. Checks conducted with distributions of 10,000 graphs do not change our results to the accuracy of our figures.  The renormalization-group trajectories of these distributions, effected by 5,000 local renormalization-group transformations at each rescaling, are followed to the sinks \cite{BerkerW} that characterize the thermodynamic phases (Table I).  Calculationally, the renormalization-group trajectories starting in a given phase unmistakably reach their respective sinks within 8 renormalization-group iterations, including the case of the nematic and spin-glass phases.  Renormalization-group trajectories starting close to a phase boundary spend iterations close to the unstable fixed point controlling the phase boundary, but reach the sink within 15 iterations for the accuracy of our calculation.  The calculational uncertainty in our work is smaller than the thickness of all of the phase boundary lines in our figures.

In each local renormalization-group transformation, the first, decimation step consists in matrix-multiplying $b=3$ transfer matrices randomly chosen from the distribution of this quenched random system.  Here $b=3$ is the length rescaling factor of the renormalization-group transformation:
\begin{equation}
\mathbf{\widetilde{T}} = \mathbf{T_1\cdot T_2\cdot T_3} \, ,
\end{equation}
The next, bond-moving step consists in multiplying the identically placed [within the matrix, namely at the same $(s_i,s_j)$ position] elements of $b^{d-1}=9$ bond-moved transfer matrices:
\begin{equation}
T'(s_i,s_j) =  \prod_{k=1}^9 \widetilde{T}_k(s_i,s_j) \, .
\end{equation}
In this equation, prime signifies renormalized, and the local renormalization-group transformation is done.

\begin{figure*}[ht!]
\centering
\includegraphics[scale=0.38]{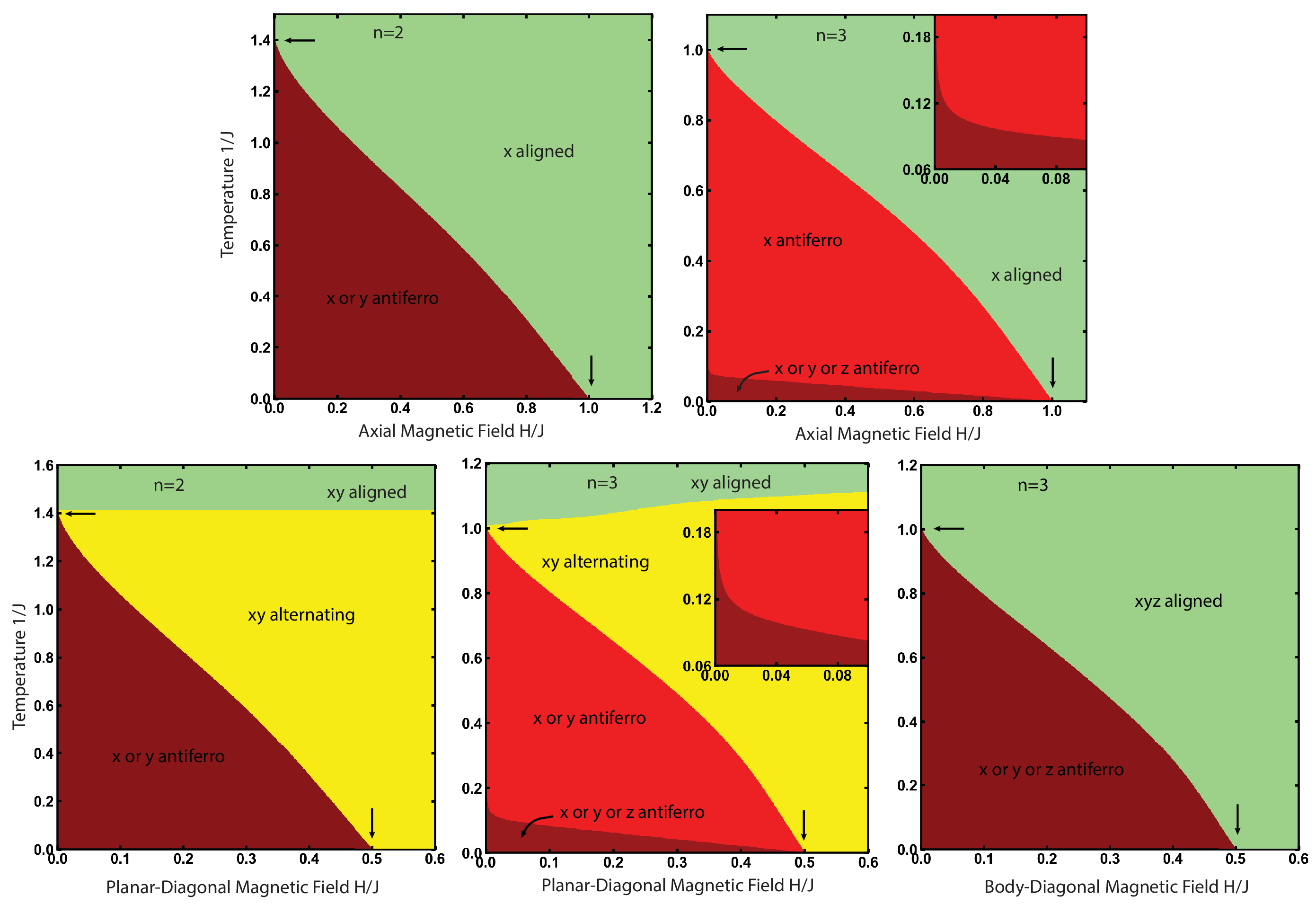}
\caption{Calculated phase diagrams for the cubic-spin system under external axial (in the $+x$ direction), planar-diagonal (in the $+xy$ direction), body-diagonal (in the $+xyz$ direction) magnetic fields, in spatial dimension $d=3$, for $p=1$, namely the antiferromagnetic system. The top row shows the application of the axial field: At high temperatures or high fields, the system aligns along the applied field.  The bottom row shows the application of planar-diagonal or body-diagonal magnetic fields: At high temperatures, the system aligns with the applied field. In all panels, at low temperatures and low fields, the system orders in the fully antiferromagnetic phase of the zero-field system.  For $n=3$ under axial or planar-diagonal magnetic fields, an intermediate, less degenerate, antiferromagnetic phase occurs in one of the directions of the axial or planar-diagonal field. In these cases, the fully antiferromagnetic phase persists asymptotically close to the zero-field axis, as seen in the insets. For planar-diagonal magnetic field, another ordered phase (doubly degenerate) of $xy$ alternation occurs and continues to all field strengths.  All finite-temperature phase boundaries meet at the transition temperature (shown with horizontal arrow) of the zero-field system, which thus gives a multiphase point \cite{Hoston} of three or four phases occurring at finite temperature.  The zero-temperature phase transitions (shown with vertical arrow) occur at the ground-state-energy crossings, which also give multiphase points of three phases.}
\end{figure*}

\begin{figure*}[ht!]
\centering
\includegraphics[scale=0.38]{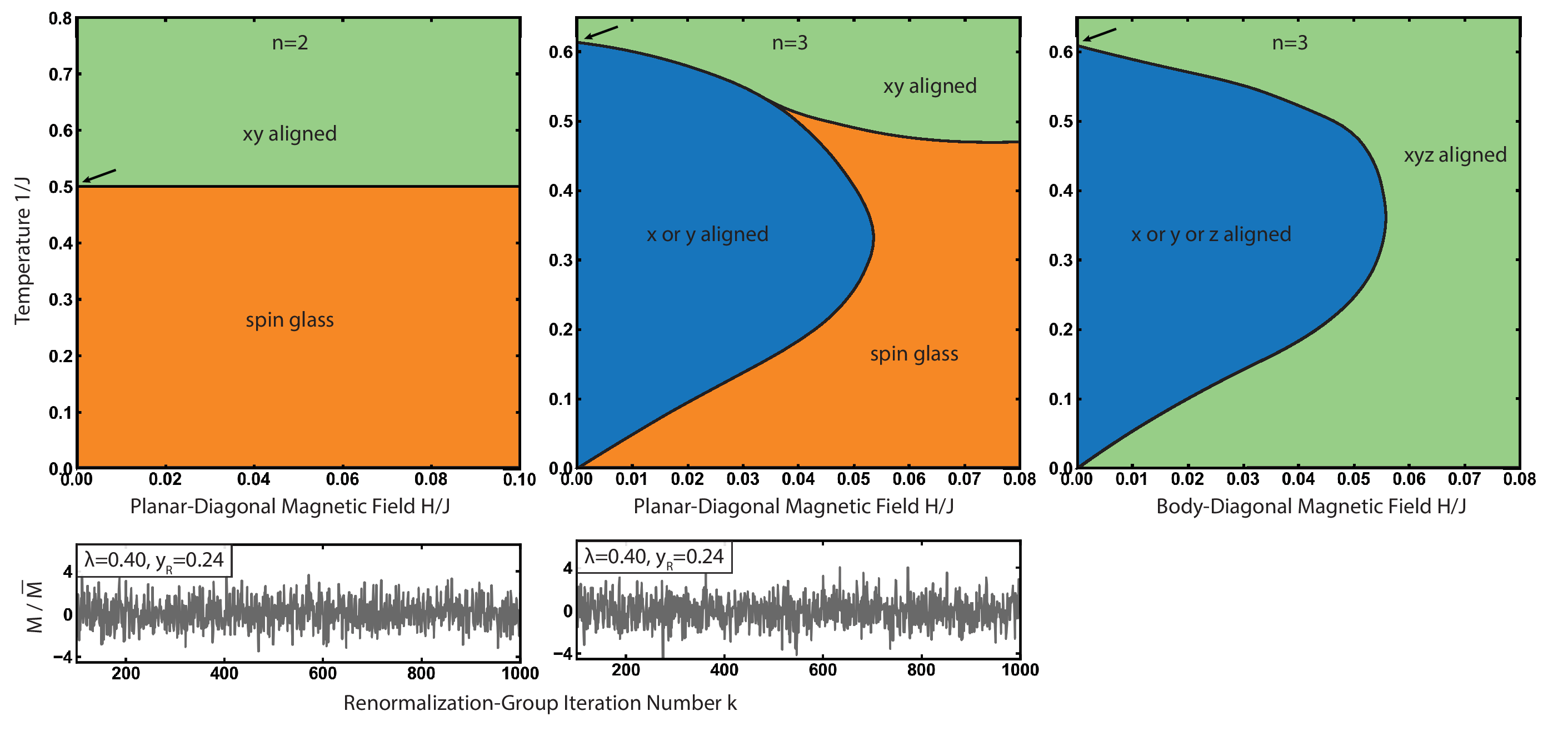}
\caption{Calculated phase diagrams for the cubic-spin system under planar-diagonal and body-diagonal magnetic fields, in spatial dimension $d=3$, for $p=0.5$, namely the spin-glass system. A spin-glass phase occurs under planar-diagonal magnetic fields.  This spin glass phase results from the asymptotic competition, under renormalization-group, of the $+x$ or $+y$ aligned phase and the $xy$ alternating phase. Both of these phases are doubly degenerate, so that the spin-glass phase is quadruply degenerate.  Thus, as shown in the second line of the figure, it is $ -\beta \mathcal{H}_{ij}(+x,+x)-\beta \mathcal{H}_{ij}(+x,+y) \equiv 2M$ that is chaotic under renormalization-group scale changes, whereas in conventional spin-glass phases it is $-\beta \mathcal{H}_{ij}(+x,+x)-\beta \mathcal{H}_{ij}(+x,-x) = 2J$ that is chaotic under renormalization-group scale changes. In the second line, these calculated chaoses, their calculated Lyapunov exponents $\lambda$ and their runaway exponents $y_R$ are given.  For both $n=2$ and 3, on the high-temperature side, the transition temperature is unaffected or weakly affected by field strength. Each of these two spin-glass phase diagram topologies here are very different from conventional spin-glass phase diagram topologies:  The axis orthogonal to temperature is not a quenched probability but the magnetic field; one of the competing phases, $xy$ alternating, does not appear in the phase diagram; the spin-glass phase stretches indefinitely in the horizontal axis direction.  In both $n=3$ cases, namely with or without the occurrence of the spin-glass phase, phase reentrances occur in the temperature direction.  The zero-field intercepts of the phase boundaries (shown with arrows) are the transition points, seen in Fig. 1, between the disordered phase and the zero-field spin-glass phase for $n=2$ and between the disordered phase and the zero-field nematic phase for $n=3$. Our (exact for hierarchical model) calculation does not give any finite-field phase boundary that ends on the zero-field spinglass-nematic transition. Thus, the nematic phase appears for field-free condition.  The renormalization-group sinks of each phase, to which all points of the phase map under renormalization-group, are given in Table I and Fig. 5.  These sinks epitomize the ordering of their respective phases.}
\end{figure*}

\section{Results: Magnetic Fields on the Ferromagnetic Phase Yield 3 Phase Diagrams}

Calculated phase diagrams for the cubic-spin system under planar-diagonal (in the $+xy$ direction) and body-diagonal (in the $+xyz$ direction) magnetic fields, in spatial dimension $d=3$, for $p=0$, namely the ferromagnetic system, are shown in Fig. 2. In all three cases, a uniaxially aligned symmetry-broken ($x$ or $y$ or $z$ aligned) ordered phase occurs at low temperatures and persists to all high fields.  A phase aligned along the applied magnetic field occurs at high temperatures.  The phase transition temperatures between these two phases are essentially independent of field strength and join, at the left intercept of the panels, the ferromagnetic transition temperature (marked by arrow) of the zero-field systems (left edge of Fig. 1). The ordered phases at finite-field are doubly $(x$ or $y$ aligned) or triply $(x$ or $y$ or $z$ aligned) degenerate. These degeneracies double at ordered phases joined at zero field, since the reverse magnetized phases also occur.

With the application to this ferromagnetic system of an axial magnetic field (in the $+x$ direction, even in infinitesimal amount), the ordered phase disappears and the system is uniaxially aligned (along $+x$) at all temperatures.

\section{Results: Magnetic Fields on the Antiferromagnetic Phase Yield 5 Phase Diagrams}

Calculated phase diagrams for the cubic-spin system under axial (in the $+x$ direction), planar-diagonal (in the $+xy$ direction), body-diagonal (in the $+xyz$ direction) magnetic fields, in spatial dimension $d=3$, for $p=1$, namely the antiferromagnetic system, are shown in Fig. 3. The top row shows the application of the axial field: At high temperatures or high fields, the system aligns along the applied field.  The bottom row shows the application of planar-diagonal or body-diagonal magnetic fields: At high temperatures, the system aligns with the applied field. In all panels, at low temperatures and low fields, the system orders in the fully antiferromagnetic phase of the zero-field system, namely antiferromagnetic in spin direction $x$ or $y$ or $z$, each doubly degenerate by spatial translation.  For $n=3$ under axial or planar-diagonal magnetic fields, an intermediate, less degenerate, antiferromagnetic phase occurs in one of the directions of the axial or planar-diagonal field. In these cases, the fully antiferromagnetic phase persists asymptotically close to the zero-field axis, as seen in the insets. For planar-diagonal magnetic field, another ordered phase (doubly degenerate by spatial translation) of $xy$ alternation occurs and continues to all field strengths.  All finite-temperature phase boundaries meet at the transition temperature (shown with horizontal arrow) of the zero-field system, which thus give a multiphase point \cite{Hoston} of three or four phases, without counting the degeneracies, occurring at finite temperature.  The zero-temperature phase transitions (shown with vertical arrow) occur at the ground-state-energy crossings, which also give multiphase points of three phases.

\begin{table*}

\begin{tabular}{c}
\multicolumn{1}{c}{Renormalization-Group Sinks of the n=2 Finite-Field Thermodynamic Phases} \\
\end{tabular}

\begin{tabular}{c c c c c c c c c  c c c c}

\hline

\vline & $\begin{pmatrix} 1 & 0 & 0 & 0 \\ 0 & 0 & 0 & 0 \\ 0 & 0 & 0 & 0 \\ 0 & 0 & 0 & 0 \end{pmatrix}$  &\vline  & $\begin{pmatrix} 1 & 0 & 1 & 0 \\ 0 & 0 & 0 & 0 \\ 1 & 0 & 1 & 0 \\ 0 & 0 & 0 & 0 \end{pmatrix}$  &\vline  &  $\begin{pmatrix} 0 & 0 & 1 & 0 \\ 0 & 0 & 0 & 0 \\ 1 & 0 & 0 & 0 \\ 0 & 0 & 0 & 0 \end{pmatrix}$  &\vline &$\begin{pmatrix} 1 & 0 & 0 & 0 \\ 0 & 0 & 0 & 0 \\ 0 & 0 & 1 & 0 \\ 0 & 0 & 0 & 0 \end{pmatrix}$ & \vline & $\begin{pmatrix} 0 & 1 & 0 & 0 \\ 1 & 0 & 0 & 0 \\ 0 & 0 & 0 & 1 \\ 0 & 0 & 1 & 0 \end{pmatrix}$ &\vline & $\begin{pmatrix} 1 & 1 & 1 & 1 \\ 1 & 1 & 1 & 1 \\ 1 & 1 & 1 & 1 \\ 1 & 1 & 1 & 1 \end{pmatrix}$ &\vline  \\
\hline
\vline & axial $x$ aligned  &\vline & planar $xy$ aligned  &\vline & $xy$ alternating (2) &\vline &$x$ or $y$ aligned (2)  &\vline & $x$ or $y$ antiferro (4) &\vline &disordered &\vline   \\
\hline

\end{tabular}

\begin{tabular}{c}
\multicolumn{1}{c}{Renormalization-Group Sinks of the n=3 Finite-Field Thermodynamic Phases} \\
\end{tabular}

\begin{tabular}{c c c c c c c c c c c c c}

\hline

\vline & $\begin{pmatrix} 1 & 0 & 0 & 0 & 0 & 0 \\ 0 & 0 & 0 & 0 & 0 & 0 \\ 0 & 0 & 0 & 0 & 0 & 0 \\ 0 & 0 & 0 & 0 & 0 & 0 \\ 0 & 0 & 0 & 0 & 0 & 0 \\ 0 & 0 & 0 & 0 & 0 & 0 \end{pmatrix}$ &\vline  & $\begin{pmatrix} 1 & 0 & 1 & 0 & 0 & 0 \\ 0 & 0 & 0 & 0 & 0 & 0 \\ 1 & 0 & 1 & 0 & 0 & 0 \\ 0 & 0 & 0 & 0 & 0 & 0 \\ 0 & 0 & 0 & 0 & 0 & 0 \\ 0 & 0 & 0 & 0 & 0 & 0 \end{pmatrix}$ &\vline & $\begin{pmatrix} 1 & 0 & 1 & 0 & 1 & 0 \\ 0 & 0 & 0 & 0 & 0 & 0 \\ 1 & 0 & 1 & 0 & 1 & 0 \\ 0 & 0 & 0 & 0 & 0 & 0 \\ 1 & 0 & 1 & 0 & 1 & 0 \\ 0 & 0 & 0 & 0 & 0 & 0 \end{pmatrix}$ &\vline &$\begin{pmatrix} 0 & 0 & 1 & 0 & 0 & 0 \\ 0 & 0 & 0 & 0 & 0 & 0 \\ 1 & 0 & 0 & 0 & 0 & 0 \\ 0 & 0 & 0 & 0 & 0 & 0 \\ 0 & 0 & 0 & 0 & 0 & 0 \\ 0 & 0 & 0 & 0 & 0 & 0 \end{pmatrix}$ &\vline &$\begin{pmatrix} 1 & 1 & 1 & 1 & 1 & 1 \\ 1 & 1 & 1 & 1 & 1 & 1 \\ 1 & 1 & 1 & 1 & 1 & 1 \\ 1 & 1 & 1 & 1& 1 & 1 \\ 1 & 1 & 1 & 1 & 1 & 1 \\ 1 & 1 & 1 & 1 & 1 & 1 \end{pmatrix}$ &\vline

\\
\hline
\vline & axial $x$ aligned &\vline & planar $xy$ aligned &\vline & body $xyz$ aligned &\vline & $xy$ alternating (2) &\vline &disordered& \vline \\
\hline

\vline &$\begin{pmatrix} 1 & 0 & 0 & 0 & 0 & 0 \\ 0 & 0 & 0 & 0 & 0 & 0 \\ 0 & 0 & 1 & 0 & 0 & 0 \\ 0 & 0 & 0 & 0 & 0 & 0 \\ 0 & 0 & 0 & 0 & 0 & 0 \\ 0 & 0 & 0 & 0 & 0 & 0 \end{pmatrix}$ &\vline  &$\begin{pmatrix} 1 & 0 & 0 & 0 & 0 & 0 \\ 0 & 0 & 0 & 0 & 0 & 0 \\ 0 & 0 & 1 & 0 & 0 & 0 \\ 0 & 0 & 0 & 0 & 0 & 0 \\ 0 & 0 & 0 & 0 & 1 & 0 \\ 0 & 0 & 0 & 0 & 0 & 0 \end{pmatrix}$ &\vline &$\begin{pmatrix} 0 & 1 & 0 & 0 & 0 & 0 \\ 1 & 0 & 0 & 0 & 0 & 0 \\ 0 & 0 & 0 & 0 & 0 & 0 \\ 0 & 0 & 0 & 0 & 0 & 0 \\ 0 & 0 & 0 & 0 & 0 & 0 \\ 0 & 0 & 0 & 0 & 0 & 0 \end{pmatrix}$ &\vline &
$\begin{pmatrix} 0 & 1 & 0 & 0 & 0 & 0 \\ 1 & 0 & 0 & 0 & 0 & 0 \\ 0 & 0 & 0 & 1 & 0 & 0 \\ 0 & 0 & 1 & 0 & 0 & 0 \\ 0 & 0 & 0 & 0 & 0 & 0 \\ 0 & 0 & 0 & 0 & 0 & 0 \end{pmatrix}$ &\vline & $\begin{pmatrix} 0 & 1 & 0 & 0 & 0 & 0 \\ 1 & 0 & 0 & 0 & 0 & 0 \\ 0 & 0 & 0 & 1 & 0 & 0 \\ 0 & 0 & 1 & 0 & 0 & 0 \\ 0 & 0 & 0 & 0 & 0 & 1 \\ 0 & 0 & 0 & 0 & 1 & 0 \end{pmatrix}$ & \vline \\
\hline
\vline &$x$ or $y$ aligned (2) &\vline & $x$ or $y$ or $z$ aligned (3)&\vline & $x$ antiferro (2)  &\vline & $x$ or $y$ antiferro (4) &\vline & $x$ or $y$ or $z$ antiferro (6) &\vline \\

\hline

\end{tabular}

\caption{Under repeated renormalization-group transformations, the phase diagram points of the phases of the finite-field $n$-component cubic-spin spin glass flow to the sinks shown on this Table (and in Fig. 5), giving the exponentiated nearest-neighbor Hamiltonians, namely the transfer matrices. The number of coexisting phases are shown in parenthesis.}
\end{table*}

\section{Results: Magnetic Fields on the Nematic/Spin-Glass Phase Yield 3 Phase Diagrams and Two Topologies}

Calculated phase diagrams for the cubic-spin system under planar-diagonal and body-diagonal magnetic fields, in spatial dimension $d=3$, for $p=0.5$, namely the spin-glass system, are given in Fig. 4. A spin-glass phase occurs under planar-diagonal magnetic fields.  It is seen that this spin-glass phase results from the asymptotic competition, under renormalization-group, of the $+x$ or $+y$ aligned phase and the $xy$ alternating phase. Both of these phases are doubly degenerate, so that the spin-glass phase is quadruply degenerate.  Thus, as shown in the second line of the figure, it is $-\beta \mathcal{H}_{ij}(+x,+x)-\beta \mathcal{H}_{ij}(+x,+y) \equiv 2M$ that is chaotic under renormalization-group scale change, whereas in conventional spin-glass phases it is $-\beta \mathcal{H}_{ij}(+x,+x)-\beta \mathcal{H}_{ij}(+x,-x) = 2J$ that is chaotic under renormalization-group scale change. Thus, for a cubic-spin spin-glass under planar-diagonal magnetic field, an Ising spin-glass phase is realized, from asymmetric phases $x$ or $y$ and $xy$.  For both $n=2$ and 3, on the high-temperature side, the transition temperature is unaffected or weakly affected by field strength. An identical behavior occurs in the "usual" spin glasses, where the high-temperature phase transition temperature does not vary or weakly varies in the phase diagram (Fig. 1).  Each of these two spin-glass phase diagram topologies here are very different from conventional spin-glass phase diagram topologies:  The axis orthogonal to temperature is not a quenched probability but the magnetic field; one of the competing phases, $xy$ alternating, does not appear in the phase diagram; the spin-glass phase stretches indefinitely in the horizontal axis direction.  In both $n=3$ cases, namely with or without the occurrence of the spin-glass phase, phase reentrances \cite{Cladis} occur in the temperature direction. Such phase reentrance behavior has been seen dipolar liquid crystals \cite{Netz,Garland}, molecular entropic binary liquid mixtures \cite{Walker}, oversaturatedly adsorbed surface systems \cite{Caflisch}, random-field tranverse Ising models \cite{transverse}, high-curvature (black hole) gravity \cite{Mann1, Mann2}. The zero-field intercepts of the phase boundaries in Fig. 4 are the transition points, seen in Fig. 1, between the disordered phase and the zero-field spin-glass for $n=2$ and the disordered phase and the zero-field nematic phase for $n=3$.  Our (exact for hierarchical model) calculation does not give any finite-field phase boundary that ends on the zero-field spinglass-nematic transition. Thus, the nematic phase appears for field-free condition.

The renormalization-group trajectories in the spin-glass phases are chaotic, as shown in the second line of Fig. 4.  The strength of chaos under scale change \cite{McKayChaos,McKayChaos2,BerkerMcKay,McKayChaos4} is measured by the Lyapunov exponent \cite{Collet,Hilborn},
\begin{equation}
\lambda = \lim _{n\rightarrow\infty} \frac{1}{n} \sum_{k=0}^{n-1} \ln \Big|\frac {dx_{k+1}}{dx_k}\Big|\,,
\end{equation}
where, in the current case, $x_k = M/\overline{M}$ at step $k$ of the renormalization-group trajectory, where $M=[-\beta \mathcal{H}_{ij}(+x,+x)-\beta \mathcal{H}_{ij}(+x,+y)]/2$ as defined above and $\overline{M}$ is the average of its absolute value in the quenched random distribution. The location $ij$ and the renormalized locations overlaying it are included in the summation, which readily converges. Thus, we calculate the Lyapunov exponents by discarding the first 100 renormalization-group steps (to eliminate crossover from initial conditions to asymptotic behavior) and then using the next 900 steps, shown in Fig. 4. As expected from the previous paragraph, these very different variable and topologies still give the Ising Lyapunov exponent of $\lambda = 0.40$, whereas non-Ising Lyapunov exponents do very commonly occur \cite{Artun}.

In addition to chaos, the renormalization-group trajectories show asymptotic strong-coupling behavior \cite{Demirtas},
\begin{equation}
\overline{M'} = b^{y_R}\, \overline{M}\,,
\end{equation}
where the prime denotes renormalized and $y_R >0$ is the strong-coupling runaway $(\overline{M'} > \overline{M})$ exponent \cite{Demirtas}.  Again using 900 renormalization-group steps after discarding 100 steps, we find here the same value of $y_R = 0.24$, which appears to be common to a large number of otherwise different spin glasses, also reflecting that spin-glass order is very unsaturated order.\cite{Artun}  The fact that $y_R$ is much less than $d-1=2$ indeed shows that this spin-glass order is very unsaturated order: In the compact (saturating) ordering of systems without quenched randomness, $y_R = d-1$ reflects the energy of the smooth interface between oppositely aligned renormalization-group cells.\cite{BerkerW}

On the other hand, with the application to this spin-glass system of an axial magnetic field (in the $+x$ direction, even in infinitesimal amount), the spin-glass phase disappears and the system is uniaxially aligned (along $+x$) at all temperatures.

\begin{figure}[ht!]
\centering
\includegraphics[scale=0.27]{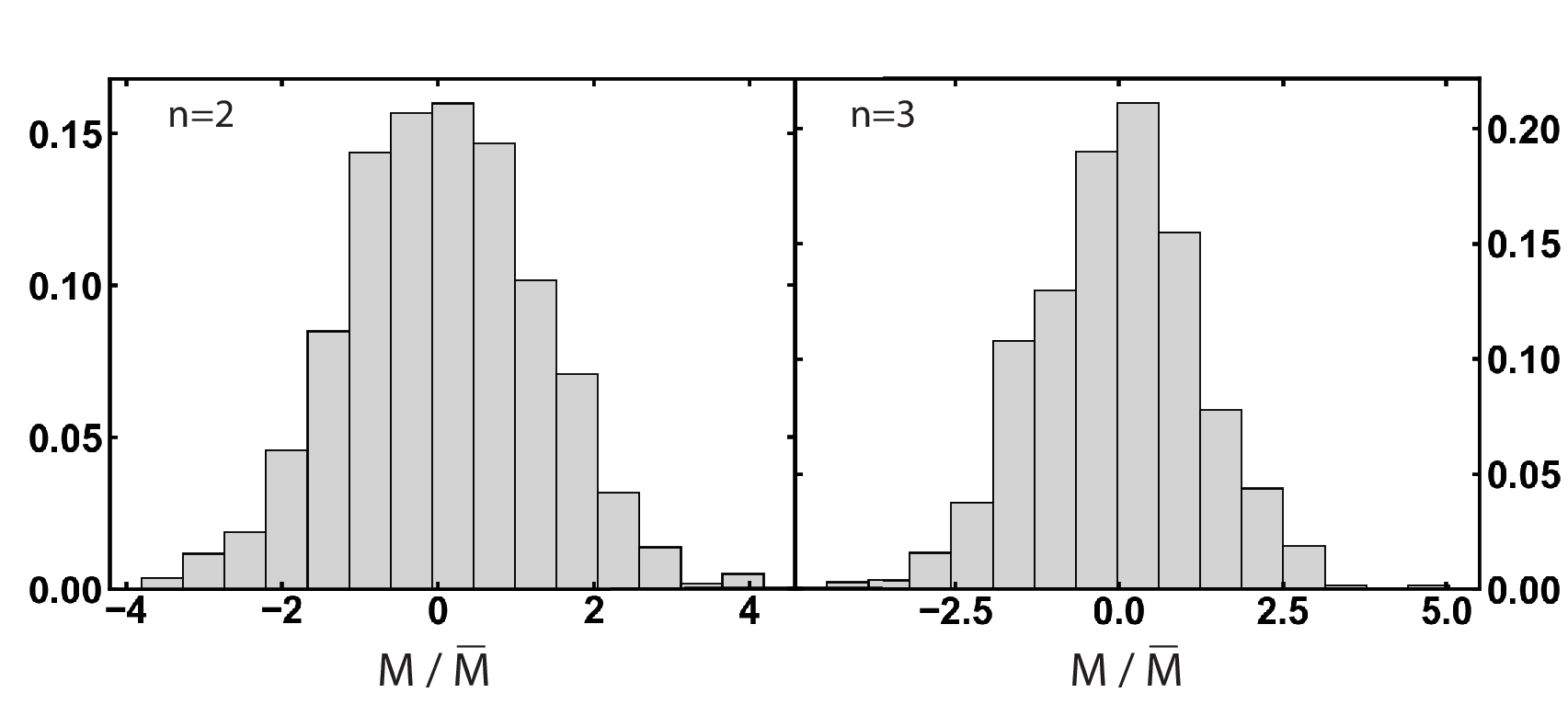}
\caption{Fixed distributions \cite{Gurleyen} that are the sinks of the spin-glass phases.  The fixed distributions of the chaotic variable $ M = [-\beta \mathcal{H}_{ij}(+x,+x)-\beta \mathcal{H}_{ij}(+x,+y)]/2$ are given. $\overline{M}$ is the average, in the distribution, of the absolute value of $M$.}
\end{figure}

\section{Conclusions}

We have solved the $n=2$ and 3-component cubic-spin spin-glass system under uniform axial, planar-diagonal, and body-diagonal magnetic fields.  We find 17 different phases including a spin-glass phase and two spin-glass phase-diagram topologies very different from the conventional spin-glass phase-diagram topologies.  It is of interest that such a discrete-spin model houses two different spin-glass phases (counting the one at zero fields) and the nematic phase (at zero field), both of which are high built-in entropy phases, the spin glass due to frustration and the nematic due to bidirectional ordering.  Furthermore, the spin-glass phase here under finite field presents two qualitatively different phase diagram topologies, namely field-continued (from zero field) and field-induced (at finite field).  Moreover, the competing ordering (e.g., ferromagnetic versus antiferromagnetic, or left versus right chiral \cite{Caglar2}, in the common spin glasses) are here $+x$ or $+y$ aligned versus $xy$ alternating, not mapping onto each other under a symmetry. The axis orthogonal to temperature is not a quenched probability but an applied magnetic field; one of the competing orderings, $xy$ alternating, does not appear as a phase in the phase diagram; the spin-glass phase stretches indefinitely along the horizontal axis, namely magnetic field direction.  These are strong departures from usual spin glasses.  On the other hand, the chaos determination in the form of the calculated Lyapunov exponent and the nonsaturated ordering seen from the positive but small strong-coupling exponent, match the usual spin glass.

In addition to the spin glasses, the model shows 14 different ordered phases.  Thus, a simple discrete-spin shows rich ordering behavior.

\begin{acknowledgments}
Support by the Kadir Has University Doctoral Studies Scholarship Fund and by the Academy of Sciences of Turkey (T\"UBA) is gratefully acknowledged.
\end{acknowledgments}

\end{document}